\documentclass[12pt]{article}

\newcommand\nice[1]{#1}    \newcommand\subm[1]{}   
\subm{ \renewcommand\dosingle[1]{ }    }

\usepackage{graphicx}

\usepackage{amsfonts}
\usepackage{mathpi}

\usepackage[numbers]{natbib}  
\usepackage{url} 

\usepackage{hyperref}

\usepackage[symbol]{footmisc}

\newcommand\mystamp[1]{#1}

\newcommand\mystamppreamble{
  \usepackage{eso-pic}
  \usepackage{color} 
  \definecolor{redstamp}{rgb}{0.99,0.80,0.90} 
  \usepackage{datetime}
  \usepackage[normalem]{ulem}
}

\mystamp{\mystamppreamble}

\providecommand{\eprint}[1]{\href{http://arxiv.org/abs/#1}{[arXiv:#1]}}

\providecommand{\url}[1]{\href{#1}{#1}}

\providecommand{\adsurl}[1]{} 

\newcommand\BIBAABST{iop_hyperref} 
\newcommand\BIBHOME{mybib} 

\providecommand\apj{Astrophys.J.}                 
\providecommand\mnras{MNRAS}

\providecommand\prd{Phys.Rev.D}

\nice{ \newcommand\mycaptionfont{\protect\footnotesize} }

\newcommand\gtapprox{\,\lower.6ex\hbox{$\buildrel >\over \sim$} \, }
\newcommand\ltapprox{\,\lower.6ex\hbox{$\buildrel <\over \sim$} \, }
\newcommand\propapprox{\,\lower.6ex\hbox{$\buildrel \propto\over \sim$} \, }

\newcommand\arcs{\ifmmode {'' }\else $'' $\fi}     
\newcommand\arcm{\ifmmode {' }\else $' $\fi}       
\newcommand\ddeg{\ifmmode^\circ\else$^\circ$\fi}    

\newcommand\frtoday{Le\space\number\day\space\ifcase\month\or
  janvier\or f\'evrier\or mars\or avril\or mai\or juin\or
  juillet\or ao\^ut\or septembre\or octobre\or novembre\or 
d\'ecembre\fi\space \number\year}

\newcommand\hMpc{\mbox{$h^{-1}$ Mpc}}

\newcommand\Omm{\Omega_{\mathrm m}}
\newcommand\Ommzero{\Omega_{\mathrm{m}0}}

\newcommand\Ommeff{\Omega_{\mathrm{m}}^{\mathrm{eff}}}

\newcommand\Omkeff{\Omega_{\mathrm{k}}^{\mathrm{eff}}}
\newcommand\chieff{\chi^{\mathrm{eff}}}
\newcommand\Heff{H^{\mathrm{eff}}}
\newcommand\Hbg{H_{\mathrm{bg}}}

\newcommand\RCeff{R_{\mathrm{C}}^{\mathrm{eff}}}
\newcommand\dLeff{d_{{L}}^{\mathrm{eff}}}

\newcommand\Hpeculiar{H_{\mathrm{pec}}}

\newcommand\diffd{\mathrm{d}}

\newcommand\fvir{f_{\mathrm{vir}}}
\newcommand\deltavir{\delta_{\mathrm{vir}}}

\begin{document}

\title{Dark energy with rigid voids versus relativistic voids alone}

\author{Boudewijn F. Roukema$^{1,2}$ \\
{\em $^1$Toru\'n Centre for Astronomy}\\
{\em Faculty of Physics, Astronomy and Informatics}\\
{\em Nicolaus Copernicus University} \\
{\em ul. Gagarina 11, 87-100 Toru\'n, Poland} \\
{\em $^2$Universit\'e de Lyon, Observatoire de Lyon}\\
{\em Centre de Recherche Astrophysique de Lyon}\\
{\em CNRS UMR 5574: Universit\'e Lyon~1 and \'Ecole Normale Sup\'erieure de Lyon}\\ 
{\em 9 avenue Charles Andr\'e, F--69230 Saint-Genis-Laval, France}\thanks{During visiting lectureship.}\\
{\tt boud@astro.uni.torun.pl}\\
{\ }\\{\ }\\
30 March 2013
}

\date{{\it Essay written for the Gravity Research Foundation 2013}\\
{\it Awards for Essays on Gravitation}
}

\maketitle

\abstract{The standard model of cosmology is dominated---at the
  present epoch---by dark energy. Its voids are rigid and Newtonian
  within a relativistic background. The model prevents them from
  becoming hyperbolic. Observations of rapid velocity flows out of
  voids are normally interpreted within the standard model that is
  rigid in comoving coordinates, instead of allowing the voids'
  density parameter to drop below critical and their curvature
  to become negative.  Isn't it time to advance beyond nineteenth century
  physics and relegate dark energy back to the ``no significant
  evidence'' box?}

\newcommand\fDEvsvirial{
  \begin{figure}  
    \centering
    \includegraphics[width=10cm]{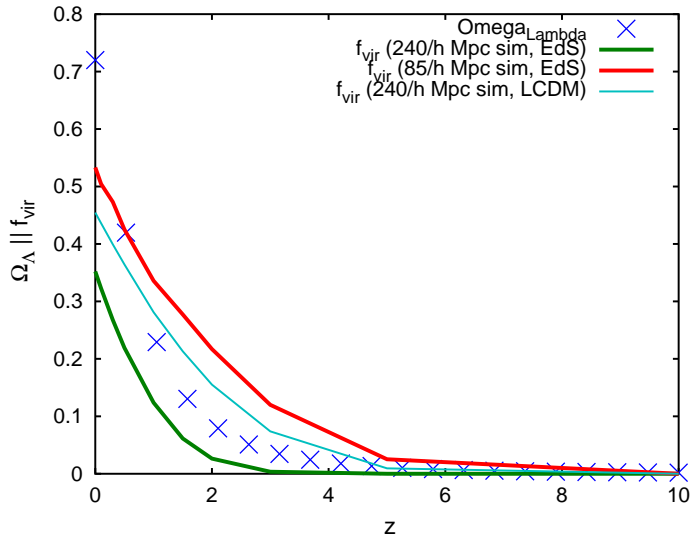}
    \caption[]{ \mycaptionfont Virialisation fraction $\fvir(z)$ in
      Virgo Consortium $256^3$-particle simulations
      \protect\citep{Jenkins98Virgo,Thomas98Virgo} with 240~{\hMpc}
      and 85~{\hMpc} box sizes, shown as continuous thick curves for
      the Einstein-de~Sitter model, and dark energy parameter
      $\Omega_\Lambda(z)$ evolution for $\Omega_{\Lambda0} = 0.72$.
      Thin curve: 240~{\hMpc} $\Lambda$CDM simulation.
      \label{f-DEvsvirial}
    }
  \end{figure}
} 

\newcommand\flumdist{
  \begin{figure}  
    \centering
    \includegraphics[width=10cm]{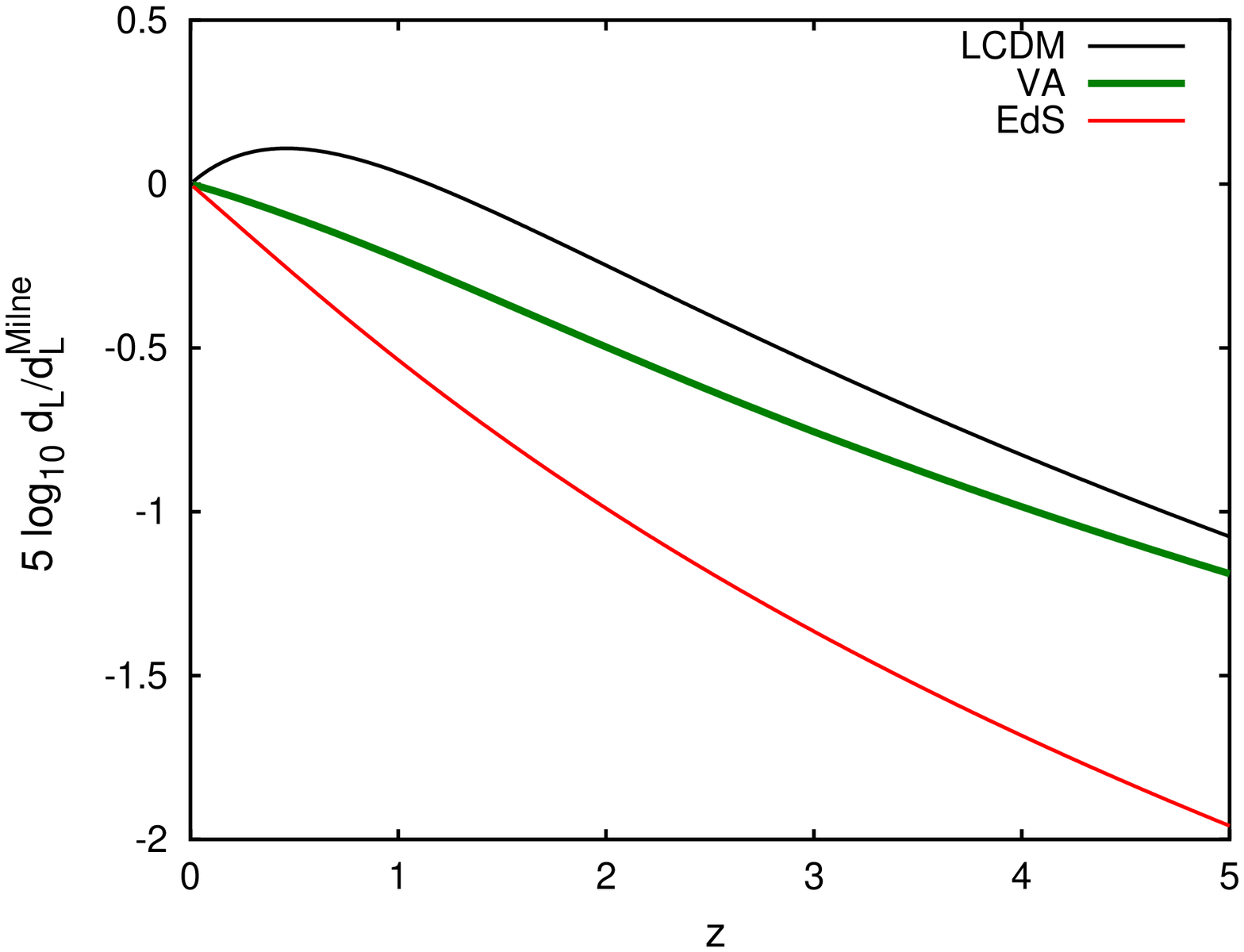}
    \caption[]{ \mycaptionfont Distance modulus normalised to the
      Milne model ($\Omm=0, \Omega_{\Lambda} = 0, \, \forall z$)
      for the homogeneous $\Lambda$CDM model (top, black), the
      uncorrected, homogeneous EdS model (bottom, red), and the
      void-corrected EdS model (middle, thick, green; ``VA'' = virialisation
      approximation).
      \label{f-lumdist}
    }
  \end{figure}
} 

\clearpage



\section{``Dark energy'' traces inhomogeneity}

\fDEvsvirial

The standard model of cosmology is generally accepted to be a spacetime
with a Friedmann--Lema\^{\i}tre--Robertson--Walker (FLRW) metric
\cite{deSitt17,Fried23,Fried24,Lemaitre31ell,Rob35}, solving the
Einstein equation rather simply thanks to the assumption of homogeneous 
density on any spatial slice, and measured to have the Concordance Model
\citep{CosConcord95} values of the matter density and dark energy 
parameters, $\Ommzero \approx 0.32$ 
(e.g. \citep{PlanckXVIcosmoparam13})
and $\Omega_{\Lambda0} := 1- \Ommzero$,
respectively (hereafter, $\Lambda$CDM).
However, we live in the inhomogeneous epoch: galaxies and voids certainly exist today.
As found by \citep{ROB13}, in an Einstein-de~Sitter (EdS) FLRW model
(i.e. with $\Ommzero=1, \Omega_{\Lambda0}=0$), 
the fraction of matter in a large region that is virialised, $\fvir$,
evolves in a very similar way to that of the dark energy parameter
in a flat FLRW model with negligible radiation density,
\begin{equation}
  \Omega_\Lambda(z) = 1 - \frac{\Ommzero}{a^3 \Omega_{\Lambda0} + \Ommzero}.
\end{equation}
This is shown in Fig.~\ref{f-DEvsvirial}, where  $\Omega_{\Lambda0} = 0.68$. 

This seems like an extraordinary coincidence. Over the same redshift
range during which one {\em expects} that the Universe is
inhomogeneous, the degree of inhomogeneity, as expressed by $\fvir(z)$
in an EdS model, approximately follows the proportion of the critical
density represented by dark energy, if the dark energy is inferred
from forcing a {\em homogeneous model} on the observational data. To
first order, $\fvir$ is not sensitive to the choice of FLRW model (see Fig.~\ref{f-DEvsvirial}), so
the coincidence also exists for inhomogeneity in a $\Lambda$CDM model.  The simplest 
inference is that a homogeneous-model--inferred non-zero dark energy parameter is 
really just {\em a measurement of inhomogeneity}.

What physical link could there be between this inhomogeneity parameter 
and homogeneous-model--inferred ``dark energy''?

\section{Void dominance: low matter density, high critical density}

The most obvious physical link between inhomogeneity and
homogeneous-inferred dark energy is the volume dominance of voids
compared to virialised regions at recent epochs
\citep{BuchCarf08,rasanenFOCUS,Buchert11Towards,buchertrasanen,wiltshireFOCUS,BuchRZA2,Larena09template,RoyBuch10chaplygin,Clarkson11backreaction,WiegBuch10}.
This is because gravitational collapse implies an increase in density,
i.e., a reduction in volume, by a factor of about $\deltavir \sim 100$--200 (e.g.,
8--18$\pi^2$ \citep{LaceyCole93MN}), so that, to first order, the collapsed matter 
occupies a negligible fraction of the spatial volume. Thus, recent-epoch
spatial volume is overwhelmingly dominated by low-density regions.

Moreover, there are velocity flows out of the voids, since otherwise,
the voids couldn't be nearly empty.  Thus, the critical density
defining spatial flatness in the voids is higher than it would be in a
homogeneous calculation. The standard, FLRW approach insists that
space expands uniformly with spatially constant curvature, i.e. 
space is rigid in comoving coordinates---it is forbidden from bending under the
influence of gravity. $N$-body simulations are typically
used to study the {\em Newtonian} formation of overdense structures
and voids within this rigid, comoving background model.
However, both the low density of the voids and the velocity flow out of them
imply that the matter density parameter in the voids is sub-critical.
Thus, the voids are hyperbolic. Geometrically, this hyperbolicity 
should also be taken into account, implying an even lower matter density parameter
(conservatively, let us ignore this: the effect is small).

\section{Volume-weighted averaged metric}

These arguments can be formalised using
the volume-weighted averaging approach to modelling inhomogeneous spatial slices
\citep{BKS00,buchert:darkenergy,BuchCarf08,BuchRZA1,BuchRZA2}, in which the Friedmann
equation is generalised to (12) of \citep{Buchert11Towards}. For simplicity,
let us (i) set the dark energy term to zero, 
(ii) neglect the kinematic backreaction as much smaller than 
the curvature backreaction (see \citep{BuchRZA2} for numerical 
justification), and (iii) combine the curvature parameters into a single full 
curvature parameter as suggested in \citep{Buchert11Towards}.
Writing ``k'' instead of ``${\cal R}$'', this gives the 
domain-averaged, effective Friedmann equation
\begin{equation}
  \Omkeff(z) = 1- \Ommeff(z).
  \label{e-defn-omkeff}
\end{equation}

Along a typical, large-scale, random, spacelike or null geodesic over recent epochs, what proportions
of the geodesic lie in the emptied and virialised regions? 
The proportions at a given $z$ are, on average,
$(1 - \fvir/\deltavir) \,:\, \fvir/\deltavir$, respectively.
Given that $\deltavir \sim 100$--200 and
\begin{equation}
  0 \le \fvir \le 1,
  \label{e-fvir-limits}
\end{equation}
less than about 1\% of the geodesic falls within the virialised regions. 
Thus, by starting with a large-scale, high-redshift, 
``background'' FLRW model---in this case, 
an EdS model---an 
effective metric can be written by assuming that the virialised matter 
contributes negligibly. 
As in \citep{ROB13}, the effective expansion rate is
\begin{equation}
  \Heff(z) =  H(z) + \Hpeculiar(z),
  \label{e-defn-Heff}
\end{equation}
combining the FLRW expansion $H(z)$ with the 
peculiar velocity gradient 
(physical, not comoving) across voids, $\Hpeculiar(z)$, 
estimated numerically from an $N$-body simulation
(\citep{Jenkins98Virgo,Thomas98Virgo}, a 240~{\hMpc} box-size EdS simulation; see \citep{ROB13})
The background Hubble constant $\Hbg$ is set to make the present value
consistent with low redshift estimates \citep{Riess11H74,Freedman12H74}, i.e.
\begin{equation}
  \Hbg := \; 74 \,\mathrm{km/s/Mpc} - \Hpeculiar(0) \; \sim 50 \mathrm{km/s/Mpc}.
  \label{e-defn-H0-lowz}
\end{equation}

Thus, the loss of matter from the voids and the higher critical density
in voids both decrease the EdS background value from $\Omega_{\mathrm{m,bg}}=1$ 
to an effective value of 
\begin{eqnarray}
  \Ommeff(z) &\approx &
  ({1-\fvir})
  \;
  \left(\frac{H}{\Heff}\right)^2 \; 
  \Omm 
  \nonumber \\
  &=& 
  ({1-\fvir})
  \;
  \left(\frac{\Hbg}{\Heff}\right)^2 \; 
  \Omega_{\mathrm{m,bg}} \; a^{-3} . \nonumber \\
  \label{e-ommeff-approx}
\end{eqnarray}
The effective radius of curvature is 
\begin{equation}
  \RCeff(z) = \frac{c}{a \Heff(z) \, \sqrt{\Omkeff(z)}}.
  \label{e-defn-RCeff}
\end{equation}
The effective metric 
(cf \citep{Larena09template}) is
\begin{eqnarray}
  \diffd s^2 &=& -\diffd t^2 + 
  a^2(t) \left[
  \diffd {\chieff}^2 +
  {\RCeff}^2 \left( \sinh^2 \frac{\chieff}{\RCeff} \right)
  (\diffd \theta^2 + \cos^2 \theta \,\diffd \phi^2) \right],
  \nonumber \\
  \label{e-metric-eff}
\end{eqnarray}
where the radial comoving component is
\begin{equation}
  \diffd \chieff(z) := \frac{c}{a^2\, \Heff(z)} \, {\diffd a}.
  \label{e-dchieff}
\end{equation}

\flumdist

\section{Matter density parameter and luminosity distances}

{\em Without any attempt to fit this approximation to observational data},
apart from (\ref{e-defn-H0-lowz}) above, the correction of the EdS model
as presented above gives an effective matter density 
parameter (\ref{e-ommeff-approx}) that drops slowly from its background value
of unity at high redshift 
down to $\Ommeff = 0.27$ at the present epoch $z=0$,
remarkably close to the last two decades' local estimates of the 
matter density parameter.

The effective luminosity distance follows 
directly from the radial
comoving distance and hyperbolicity,
\begin{equation}
    \dLeff = (1+z) \RCeff \sinh \frac{\chieff}{\RCeff}.
    \label{e-dL-eff}
\end{equation}
Figure~\ref{f-lumdist} 
shows that despite the rough nature of the virialisation approximation, 
it shifts the homogeneous EdS
magnitude--redshift relation by a substantial fraction towards the homogeneous $\Lambda$CDM
relation, and thus, towards the observational supernovae type Ia relation.

\section{Conclusion}
A handful of simple formulae, lying at the heart
of homogeneous, spatially rigid cosmology, remain approximately
valid when generalised to inhomogeneous, spatially flexible cosmology
\citep{Buchert11Towards} and
applied to what observationally and theoretically dominate the 
present-day spatial volume---the voids. 
The result is a correction to a large-scale, high-redshift, 
background Einstein-de~Sitter 
cosmological model. The correction approximately gives the observed low-redshift matter density
parameter and nearly matches the type Ia supernovae luminosity 
distance relation. 
The amplitude of the correction is unlikely to be much smaller
than estimated here. At the present epoch, 
direct observations \citep{Tully08preHpec},
$N$-body simulations \citep{Jenkins98Virgo,Thomas98Virgo}, and 
the existence of the cosmic web itself establish
inhomogeneous peculiar velocity gradients of $\sim 20$--30~km/s/Mpc,
forcing, at least, a factor of $\sim (75/50)^2$ reduction of the
Einstein-de~Sitter matter density to $\Ommeff < 0.5$, via
Eqs~(\ref{e-defn-Heff}), 
(\ref{e-defn-H0-lowz}), and (\ref{e-ommeff-approx}).
A virialisation fraction
of the order of $\sim$ 50\% reduces this to $\sim 0.25$.
Even when forcing the homogeneous FLRW models onto the data, 
the initial analysis of the Planck Surveyor 
cosmic microwave background data finds 
$\Hbg = 67.3 \pm 1.2$~km/s/Mpc at $z \approx 1100$
\citep{PlanckXVIcosmoparam13}---not as low
as the velocity gradients imply, but still significantly lower
than the low redshift estimates 
of $\Heff(0) = 74.0 \pm 1.6$~km/s/Mpc 
(\citep{Freedman12H74,Riess11H74}, standard error in the mean).

How could such a simple, back-of-the-envelope calculation have been
missed for so long? 
While the volume-averaging approach to cosmology has been developed
over many years (see e.g. \citep{Buchert11Towards} for a review),
possibly the answer lies, ironically, in
confusion between the {\em spacelike, unobserved}, comoving, present time
slice and the past light cone. A gigaparsec-scale void in the former
should have a very weak ($\delta \sim 10^{-5}$) underdensity, and our
would-be location at its centre would be uncomfortably
anti-Copernican. But these are both moot points!
On the past light cone, 
an {\em average}, gigaparsec-scale, sub-critical ($0 < \Ommeff < 0.8$) void 
is perfectly natural, since it is 
defined by the onset of the virialisation epoch at $z \ltapprox 3$. 
Moreover, we are naturally located at this pseudo-void's centre,
by the nature of the past light cone.

What is simpler: relativistic, hyperbolic voids, 
observed by an observer at the tip of the past light cone, 
with no dark energy parameter?
Or rigid, Newtonian voids together with a dark energy parameter
that traces the virialisation fraction?


\bigskip

{\em Acknowledgments:}
This essay is in part based on
work conducted within the ``Lyon Institute of
Origins'' under grant ANR-10-LABX-66,
and on
Program Oblicze\'n Wielkich Wyzwa\'n nauki i techniki (POWIEW)
computational resources (grant 87) at the Pozna\'n 
Supercomputing and Networking Center (PCSS).

\bibliographystyle{\BIBAABST} 
\bibliography{\BIBHOME}


\end{document}